\documentclass[conference]{IEEEtran}
\IEEEoverridecommandlockouts
\usepackage{fontenc,algorithm,algorithmic}
\usepackage{amssymb,amsfonts}
\usepackage{amsmath, bm}
\usepackage{graphicx}
\usepackage{epsfig}
\usepackage{epstopdf}
\usepackage{textcomp}
\usepackage{float}
\usepackage{xcolor}

\begin{document}
\title{Channel Estimation and  Detection in  FBMC/OQAM System with Affine Precoding and Decoding}
\author{\IEEEauthorblockN{Radhashyam Patra*, Arunanshu Mahaptro*}
\IEEEauthorblockA{\textit{*Department of Electronics and Telecommunication Engineering, VSSUT Burla, India, 768018}}\\
}
\maketitle
\begin{abstract}
We derive the mathematical equations required for channel estimation and data detection of  filter bank multi-carrier (FBMC) offset quadrature amplitude modulation (OQAM) systems with affine precoding and decoding. The mean square error (MSE) in least square (LS) channel estimation and bit error rate (BER) of the system is found for different training power coefficients. The proposed system gives better BER performance compared to other cutting edge FBMC systems. The optimum training power coefficient $(\sigma_{c_{opt}}^2)$ is also found from the simulation results. The band width efficiency of the system is also calculated. 
\end{abstract}
\section{Introduction}
\textbf{I}mplementation of prototype filters  in filter bank multi-carrier (FBMC) offset quadrature amplitude modulation (OQAM) systems decrease
 side lobes  strength and hence the adjacent channel
interference is reduced [1]. The prototype filters employ the
 time-frequency localization feature that makes FBMC systems
immune to dispersive channels. Thus, FBMC has become as an
alternate to OFDM as it is a robust system and is also immune
to multipath fading. In the same time, FBMC produces
intrinsic imaginary interference (III). This is because FBMC supports
orthogonality to real signals only. Therefore, the channel
estimation in FBMC has become an interesting research area.
The absence of training or pilot in blind estimation methods
demand long statistics of received data to estimate the channel.
In the FBMC system based on the method of interference
approximation (IAM), first, the III is estimated approximately
from adjacent symbols. Then, the  estimation and 
detection is done [2]. The odd symmetry property
 of the prototype filters cancels the III in  interference cancellation method (ICM) method [3]. Concatenation of data and training sequence is a feature of complex training sequence decomposition
(CTSD) method, which leads to spectrum inefficiency [4].
This loss is dealt with by superimposing data on the training
sequence [5]. However, superimpose results in interfernce between data and training in estimation as well as demodulation process. This issue is addressed in this proposed method. The $\bf{contribution}$ of our work is  listed below.
\begin{itemize}
\item Necessary mathematical equations are derived for channel estimation and data detection  in FBMC/OQAM system with affine precoding and decoding. 
\item The optimum training sequence power coefficient ($\sigma_{c_{opt}}^2$) is found from the simulation results.
\item Spectrum efficiency of the proposed method is also calculated.
\end{itemize}

\textbf{Notations:}  Bold face capital letters ($\mathbf{A}$) repesents matrices and  $a_{i,j}$ represents the $i^{th}$ row and  $j^{th}$ columnn element of matrix  $\mathbf{A}$. $ \ast $, $(\cdot)^*$,  $(\cdot)^H$, $\mathbb{E}(\cdot)$,  $\mathbb{E}(||\cdot||^2)$ represents convolution, complex conjugate, Hermitian, expectation and Frobeinus norm respectively. $j=\sqrt{-1}$, $\Re\{\cdot\}$ and $\Im\{\cdot\}$ represents the real and imaginary part of a complex mumber respectively. $\mathbf{I}_N$ and $\mathbf{0}_{a\times b}$ represents identity matrix of size $N \times N$ and zero matrix of dimension $a \times b $ respectively. 

 
\section{SISO FBMC/OQAM SYSTEM MODEL}
\subsection{Transmitted Signal}
We consider FBMC/OQAM system having $ N $ sub-carriers with sub-carrier spacing of $1/T_s$. Here, $T_s$ and $T_s/2$ represents the interval of complex symbols and real  OQAM symbols respectively. The prototype filter used in this model is Bellanger's Phydas project filter having impulse response $ p[k]$, whose coefficients are symmetrical and real. The length of the filter is $ L_p$ and the energy  is unity, due to which the energy of the original signal remains unaltered  [6]. The matrix  $\mathbf{S'}\in \mathbb{R}^{N\times K/2}$ contains the raw data elements which are independent and identically distributed (i.i.d.) having zero mean and unit variance. The elemnents of  $\mathbf{S'}$ are  QPSK modulated and their complex constellation points are stored in matrix $\mathbf{S}\in \mathbb{C}^{N\times K/2}$ at their respective locations. The real and imaginary parts of elements of $\mathbf{S}$  are extracted  according to OQAM concept and are stored in $\mathbf{X}\in \mathbb{R}^{N \times K}$ as given  below.
\begin{equation}\label{A}
\begin{split}
x_{m,2n}=
\begin{cases}
\Re(s_{m,n}),\, \, \, \, \,  m\, \, \, \text{even}
\\
\Im(s_{m,n}),\, \, \, \, \,  m\, \, \, \text{odd}
\end{cases}\\
x_{m,2n+1}=
\begin{cases}
\Im(s_{m,n}),\, \, \, \, \,  m\, \, \, \text{even}
\\
\Re(s_{m,n}),\, \, \, \, \,  m\, \, \, \text{odd}
\end{cases}
\end{split},
\end{equation}
where the sub-carrier index $m=0,\cdots ,N-1 $   and the OQAM symbol index $ n= 0,\cdots,K/2-1$. The precoder matrix $\mathbf{P}\in \mathbb{R}^{K\times (K+n)}$, training sequence matrix $\mathbf{C}\in \mathbb{R}^{N\times(K+n)}$, estimator matrix $\mathbf{E}\in \mathbb{R}^{(K+n)\times N} $ and detector matrix $\mathbf{D}\in \mathbb{R}^{(K+n)\times K} $ are designed using orthogonal matrix $ \mathbf{\Phi} \in \mathbb{R}^{(K+n)\times(K+n)}$ as given in [7]. 
\begin{eqnarray}\label{W}
\begin{split}
\mathbf{P}=\sqrt{(K+n)/K} \mathbf{\Phi}((N+1): (N+K) , :)\\
\mathbf{E}=\mathbf{\Phi}(1:N,:)^H \\
\mathbf{C}=\sqrt{K+n}.\mathbf{\Phi}(1:N,:)\\
\mathbf{D}=\mathbf{P}^H(\mathbf{P}\mathbf{P}^H)^{-1}
\end{split}\,\, , 
\end{eqnarray}
where $ n$ is a number such that $ n\geq N $.  The above matrices are designed so as to satisfy conditions given below.
\begin{equation}\label{zz}
\mathbf{PD=I}_K,\,\mathbf{PE=0}_{K \times N},\,\mathbf{CE=I}_N \,\text{and}\,\mathbf{CD=0}_{N \times K}.
\end{equation}

The covariance matrices of $ \mathbf{ C} $, $ \mathbf{ P} $, $ \mathbf{ E} $ and $ \mathbf{ D} $ can be obtained as follows: 
\begin{eqnarray}
\begin{split}
\mathbb{E}\{\mathbf{CC}^H\}=(K+n)\mathbf{I}_N,\,\, \mathbb{E}\{\mathbf{PP}^H\}=(K+n)/K \mathbf{I}_K,\\ 
\mathbb{E}\{\mathbf{E}^H\mathbf{E}\}=1/(K+n) \mathbf{I}_N,\,\, \mathbb{E}\{\mathbf{D}^H\mathbf{D}\}=K/(K+n) \mathbf{I}_K.
\end{split}
\end{eqnarray} 

Now, matrix $\mathbf{X}$ is post multiplied with precoder matrix $\mathbf{P}$ and then the training matrix $\mathbf{C}$ is superimposed  to get affine precoded matrix  $\mathbf{Z}\in \mathbb{R}^{N \times (K+n)}$ as given below.
\begin{equation}\label{1}
\mathbf{Z}=\sigma_s \mathbf{XP}+\sigma_c\mathbf{ C},
\end{equation}
where $\sigma_c^2$ is the power of training symbols and  $\sigma_s^2$ is the data power coefficient such that  $\sigma_s^2+ \sigma_c^2=1$.  In [7] it  is proved that the variance of each element of $\mathbf{X}$ and $\mathbf{XP}$ remains same i.e. unity. Therefore, the discrete time FBMC signal at the  $ k^{th}$ sample index is represented as [1]
\begin{equation}\label{B}
 s[k]=\sum\limits_{m=0}^{N-1}\sum_{n\in{\mathbb{Z}}}z_{m,n} \chi_{m,n}[k]\,\,\, ,
\end{equation}  
where we can define the FBMC basis function $\chi_{m,n}[k]$ as 
$\chi_{m,n}[k] =e^{ j\phi _{m,n}}e^{\frac{j2\pi mk}{N}} p[k-n\frac{N}{2}]$.
The prototype filter $p[k]$ used is symmetrical, real valued and is of length $N_p=bN$ where the integer $b$  is known as overlapping factor. Phase factor is defined as $\phi_{m,n}=\pi/2 (m+n)-\pi mn$. 
\subsection{Received Signal}
We consider frequency seletive time invariant wireless channel having $L_h$ tap length  and impulse response $h[k]= [h_1 \,\,\, h_2 \cdots  h_{L_h}]$. The received signal $y[k]$ is 
\begin{eqnarray}\label{G}
y[k] &=& s[k]*h[k]+\eta[k]\nonumber\\
&=&\sum\limits_{q=0}^{{L_h}-1}h[q]\sum\limits_{m=0}^{N-1}\sum_{n\in{\mathbb{Z}}}z_{m,n}\chi_{m,n} [k-q]+\eta[k] \,\,\,\, ,
\end{eqnarray}
where $\eta[k]$ represents  additive white Gaussian noise (AWGN) having zero mean and  variance $\sigma_{\eta}^2$. After FBMC demodulation, the received signal $y_{\bar{m},\bar{n}}$ at the $\bar{m}^{th}$ sub-carrier  and $\bar{n}^{th}$  OQAM symbol time   is given by
\begin{equation}\label{H}
y_{\bar{m},\bar{n}}=\sum\limits_{k=0}^{\infty} y[k]\chi^{*}_{\bar{m},\bar{n}}[k]=
\sum\limits_{m=0}^{N-1}\sum_{n\in{\mathbb{Z}}}z_{m,n}H_m  \xi_{m,n}^{\bar{m}, \bar{n}}+{\eta}_{\bar{m}, \bar{n}},\,\,
\end{equation}
where   $  H_m=\sum\limits_{q=0}^{{L_h}-1}h[q]e^{\dfrac{-j2\pi mq}{N}}$ is the $ m^{th}$ element of $ N $ point DFT of $h[q]$,\,\, ${\eta}_{\bar{m}, \bar{n}}=\sum\limits_{k=0}^{\infty}\eta[k]\chi^{*}_{\bar{m}, \bar{n}}[k]$  and    $\xi_{m,n}^{\bar{m}, \bar{n}}=\sum\limits_{k=0}^{\infty}\chi_{m,n}[k] \chi_{\bar{m}, \bar{n}}^*[k]$. It is known that [1]
\begin{equation}\label{AA}
\xi_{m,n}^{  \bar{m}, \bar{n}}=
\begin{cases}
1 , \,\,\,\,\,\,\,\,\,\,\,\,\,\,\,\,\,\,\,\, if\,\, (m,n)=(\bar{m},\bar{n})
\\
j\langle \xi \rangle _{m,n }^{ \bar{m}, \bar{n}},\,\,\,  if \,\,(m,n)\neq(\bar{m},\bar{n})
\end{cases} \, ,
\end{equation}
where $\langle \xi \rangle _{m,n }^{ \bar{m}, \bar{n}}=\Im \left\{\sum\limits_{k=0}^{\infty}\chi_{m,n}[k] \chi_{\bar{m}, \bar{n}}^*[k]\right\}$ represents the imaginary part of the trans-multiplexer. Therefore,  ($\ref{H}$) can be re-written as
\begin{eqnarray}\label{I}
y_{\bar{m}, \bar{n}}&=&H_{\bar{m}}z_{\bar{m}, \bar{n}}\xi_{\bar{m}, \bar{n}}^{\bar{m}, \bar{n}}+\sum\limits_{\substack{m=0,\\m\neq \bar{m}}}^{N-1}\sum_{\substack{n\in{\mathbb{Z}},\\n\neq \bar{n}}}H_mz_{m,n}\xi_{m,n}^{\bar{m}, \bar{n}} +{\eta}_{\bar{m}, \bar{n}}\nonumber\\
&=& H_{\bar{m}}z_{\bar{m}, \bar{n}}+j\sum\limits_{\substack {(m,n)\neq \\ (\bar{m}, \bar{n})}}H_mz_{m,n}\langle\xi\rangle _{m,n}^{\bar{m}, \bar{n}}+{\eta}_{\bar{m}, \bar{n}}.
\end{eqnarray}
Without loss of generality it can be assumed that with a well localized prototype filter in time-frequency domain, the III is contributed from first order neighborhood of ($\bar{m},\bar{n}$) i.e. $\Omega_{\bar{m},\bar{n}}=\left\{(\bar{m} \pm 1, \bar{n} \pm 1),(\bar{m} \pm 1, \bar{n}),(\bar{m} , \bar{n} \pm 1) \right\}$. Assuming the channel frequency response (CFR) to be constant over this neighborhood, it can be written as
\begin{equation}\label{BB}
y_{\bar{m}, \bar{n}}\approx H_{\bar{m}}\left\{ z_{\bar{m}, \bar{n}} +j\underbrace{\sum\limits_{(m,n)\in \Omega_{\bar{m},\bar{n}}}z_{m,n}\langle\xi\rangle _{m,n}^{\bar{m}, \bar{n}}}_\text{III} \right\}    +{\eta}_{\bar{m}, \bar{n}}
\end{equation} 
Here III represents the intrinsic imaginary interfernce which arises due to both precoded data ($\sigma_s\mathbf{XP}$) as well as superimposed training sequence ($\sigma_c\mathbf{C}$) whose contributions in matrix form can be denoted as $\mathbf{\Theta}_1$ and $\mathbf{\Theta}_2$ respectively. The received samples $ y_{\bar{m}, \bar{n}} $  in ($\ref{BB}$) can be represented in matrix form $\mathbf{Y}\in \mathbb{C}^{N \times (K+n)}$  as
\begin{equation}\label{M}
\mathbf{Y}=\mathbf{H}\{\mathbf{Z}+j(\mathbf{\Theta}_1+\mathbf{\Theta}_2)\}+\mathbf{W} ,
\end{equation}
where  $\mathbf{H}\in \mathbb{C}^{N \times N}$ is a diagonal matrix containing elements $\{H_0, H_1, \cdots , H_{N-1}\}$. Using (\ref{1}) we get 
\begin{eqnarray}\label{O}
\mathbf{Y}=\sigma_s \mathbf{H}\mathbf{XP}+\sigma_c \mathbf{H}\mathbf{C}+j\mathbf{H}(\mathbf{\Theta}_1+\mathbf{\Theta}_2)+\mathbf{W}.
\end{eqnarray}
\subsection{Channel Estimation}
Using the received signal matrix $\mathbf{Y}$, the performance of the channel estimator will be poor. This is because, the estimator will be affected by  $\sigma_s \mathbf{H}\mathbf{XP}$. Therefore, to nullify this effect estimator matrix $\mathbf{E}$ is post-multiplied with ($\ref{O}$). In [7] it is proved that post multiplication of $\mathbf{E}$ does not amplify the power of the signal. Now, the modified received signal matrix is
\begin{eqnarray}\label{P}
\mathbf{YE}&=&\sigma_s \mathbf{HXPE}+\sigma_c \mathbf{HCE}+j\mathbf{H}(\mathbf{\Theta}_1+\mathbf{\Theta}_2)\mathbf{E}+\mathbf{WE}\nonumber\\
&=& \sigma_c \mathbf{H}+j\mathbf{H}(\mathbf{\Theta}_1+\mathbf{\Theta}_2)\mathbf{E}+\mathbf{WE}\,.
\end{eqnarray}
The second term in (\ref{P}) is due to intrinsic imaginary interference from precoded data as well as superimposed training which is not present in OFDM. So the channel estimation in FBMC has become challenging. In conventional superimposed training based FBMC systems, data and training sequence interefernce is present in channel estimation which is eliminated in affine precoding and decoding method because of orthogonality property.
Now, the least square (LS) channel estimator can be obtained as 
\begin{equation}\label{S}
\mathbf{\widehat{H}}\in \mathbb{C}^{N \times N}=\sigma_c \mathbf{H}+j\mathbf{H}(\mathbf{\Theta}_1+\mathbf{\Theta}_2)\mathbf{E}+\mathbf{WE}.
\end{equation}
where $\mathbf{\widehat{H}}$ is a diagonal matrix.  The channel estimation error matrix is $\mathbf{\check{H}}=\mathbf{H}-\mathbf{\widehat{H}}=-\{(\sigma_c-1)\mathbf{H}+j\mathbf{H}(\mathbf{\Theta}_1+\mathbf{\Theta}_2)\mathbf{E}+\mathbf{WE}\}$ and the mean squared error (MSE) of the estimator is defined as
\begin{equation}\label{aaa}
\text{MSE}=\frac{1}{N}\mathbb{E}\{||\mathbf{\check{H}}||^2\}=\frac{1}{N}\mathbb{E}\{||\mathbf{H}-\mathbf{\widehat{H}}||^2\}\,\, ,
\end{equation}
where $\mathbf{H}$ and $\mathbf{\widehat{H}}$ are  diagonal  matrices. Since $\mathbf{H}$, $(\mathbf{\Theta_1+\Theta_2})$, $\mathbf{E}$ and $\mathbf{W}$ are statistically independent matrices and $\mathbf{E}$ is orthogonal, all the three terms of (\ref{S}) are statistically independent.
\begin{figure}[t!]
  \includegraphics[width=1.1\linewidth]{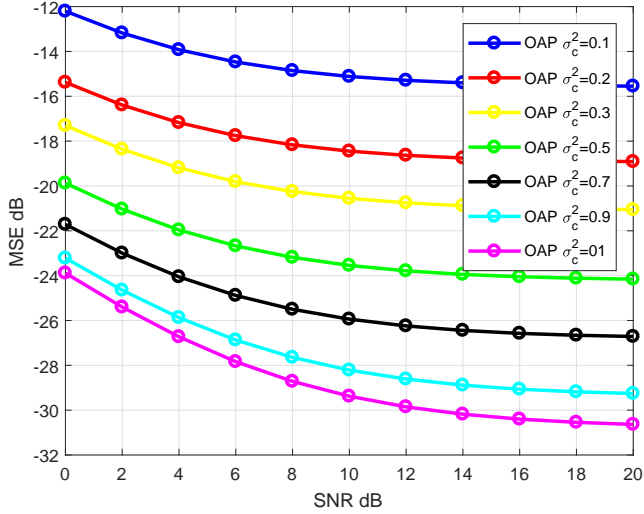}
  \caption{MSE of Channel Estimation of Proposed System $(n=N)$.}
  \label{fig:Capture4}
\end{figure}
\subsection{Data Detection}
For detection of data,  the interference by superimposed training matrix $\mathbf{C}$ have to be cancelled  by post multiplying  the detector matrix $ \mathbf{D} $ with the received signal matrix $\mathbf{Y}$  given in ($\ref{O}$). In [7] it is proved that post multiplication of $\mathbf{D}$ does not amplify the power level of the signal. The modified received signal used to detect the transmitted data is 
\begin{eqnarray}\label{T}
\mathbf{YD}&=&\sigma_s \mathbf{HXPD}+\sigma_c \mathbf{HCD}+j\mathbf{H}(\mathbf{\Theta}_1+\mathbf{\Theta}_2)\mathbf{D}+\mathbf{WD}\nonumber\\
&=&\sigma_s \mathbf{HX}+j\mathbf{H}(\mathbf{\Theta}_1+\mathbf{\Theta}_2)\mathbf{D}+\mathbf{WD},
\end{eqnarray} where (\ref{T}) follows (\ref{zz}).
The OQAM symbols are detected by pre-multiplying  ($\ref{T}$) with ${{\mathbf{\widehat{H}}}}^{-1}$ and taking the real part only as given below.
\begin{eqnarray}\label{U}
\mathbf{\widehat{X}}&=& \Re\{\mathbf{\widehat{H}}^{-1}\mathbf{YD}\}\nonumber\\
&=&\Re\{{{\mathbf{\widehat{H}}}}^{-1}\sigma_s \mathbf{HX}+j{{\mathbf{\widehat{H}}}}^{-1}\mathbf{H}(\mathbf{\Theta}_1+\mathbf{\Theta}_2)\mathbf{D}+{{\mathbf{\widehat{H}}}}^{-1}\mathbf{WD}\}\nonumber\\
&\approx & \Re\{\sigma_s \mathbf{X}+j(\mathbf{\Theta}_1   +\mathbf{\Theta}_2)\mathbf{D}+{\mathbf{\widehat{H}}}^{-1}\mathbf{WD}\}.
\end{eqnarray}
The second term in (\ref{T}) is due to intrinsic imaginary interference from precoded data as well as superimposed training which is not present in OFDM. However, its effect on data detection in FBMC is reduced after taking the real part of the signal as in (\ref{U}). In conventional superimposed training based FBMC systems, data and training sequence interefernce is present in data detection which is eliminated in affine precoding and decoding method because of orthogonality property. Rearrangement of  the elements of $\mathbf{\widehat{X}}$  to get back the complex symbols  $\mathbf{\widehat{S}}\in \mathbb{C}^{N \times K/2}$ is given by
\begin{equation}\label{V}
\hat{s}_{m,n}=
\begin{cases}
\hat{x}_{m,2n}+j\hat{x}_{m,2n+1} ,\, \, \, \, \,  m\, \, \, \text{even} 
\\
\hat{x}_{m,2n+1}+j\hat{x}_{m,2n} ,\, \, \, \, \,  m\, \, \, \text{odd}
\end{cases}.
\end{equation}
Then by taking QPSK demodulation of $\mathbf{\widehat{S}}$ we will get the actual message transmitted.
\subsection{Band Width Efficiency}
After implementation of OQAM concept, the data matrix  is $\mathbf{X}\in \mathbb{R}^{(N \times K)}$. If no precoding and decoding is applied, then to transmit $N \times K$ OQAM symbols we need $K$ symbol instants. But because of affine precoding and decoding, the  matrix becomes $\mathbf{Z}\in \mathbb{R}^{N \times (K+n)}$. So $K+n$ symbol instants are required for transmission of same data matrix. So the band width efficiency ($\text{BW}_{\text{eff}}$) of the proposed method can be defined in terms of number of raw OQAM symbols transmitted in a single affine precoded frame.  Therefore, we can write
\begin{equation}\label{zzz}
\text{BW}_{\text{eff}}=\dfrac{NK}{N(K+n)}=\dfrac{K}{K+n}.
\end{equation}
\begin{figure}[t!]
  \includegraphics[width=\linewidth]{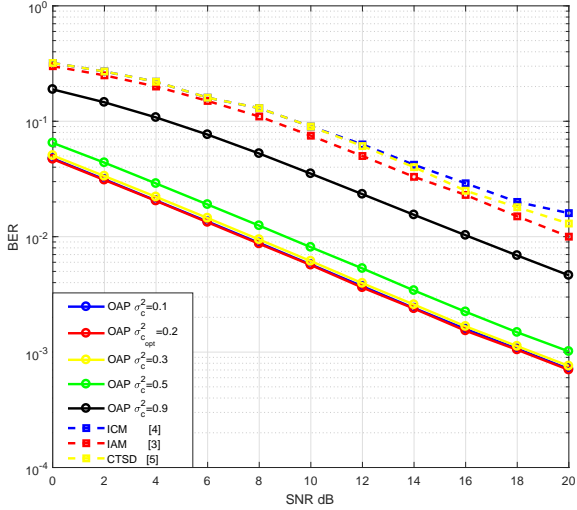}
  \caption{BER vs SNR of Propsed System $(n=N)$ with Other Systems. }
  \label{fig:Capture1}
\end{figure}
\section{SIMULATION RESULTS}
The simulations are done with the  parameters like number of sub-carriers $N  =256$, prototype filter length $L_p=4N$,  channel tap length $L_h=12$   and QPSK modulation. The real and imaginary parts of complex constellation points after QPSK modulation are separately extracted using OQAM concept. 
 The simulation is carried out  for different values of $\sigma_c^2$ and signal to noise ratio (SNR)  without channel encoding. 100 number of iterations are done to carry out the simulation. The usefulness of the proposed system is validated in terms of BER.

Fig.1  represents $(\ref{aaa})$ i.e. the MSE in LS channel estimation for different training power coefficients $\sigma_c^2={0.1, 0.2, 0.3, 0.5, 0.7, 0.9, 1.0}$. From $(\ref{aaa})$ it's clear that as $\sigma_c^2$ increases from $0$ to $1$, the error decreses which is validated in this figure. Right hand side of $(\ref{S})$ contains three terms: first one is useful for channel estimation, second is related to III and third one is noise added in channel. The second term is having significant power [8] which is not present in OFDM. 

 Fig.2  compares  BER   of the proposed  system for $n=N$ with IAM based 4x4 MIMO-FBMC system  ,  ICM based 4x4 MIMO-FBMC system and  CTSD based 4x4 MIMO-FBMC system.  An improvement of about 12dB is observed for $\sigma_c^2=\{0.1, 0.2, 0.3\}$ and  nearly 4 dB for $\sigma_c^2=\{ 0.9\}$ as compared to IAM, ICM and CTSD methods at BER of $10^{-2}$.
 \begin{figure}[t!]
  \includegraphics[width=\linewidth]{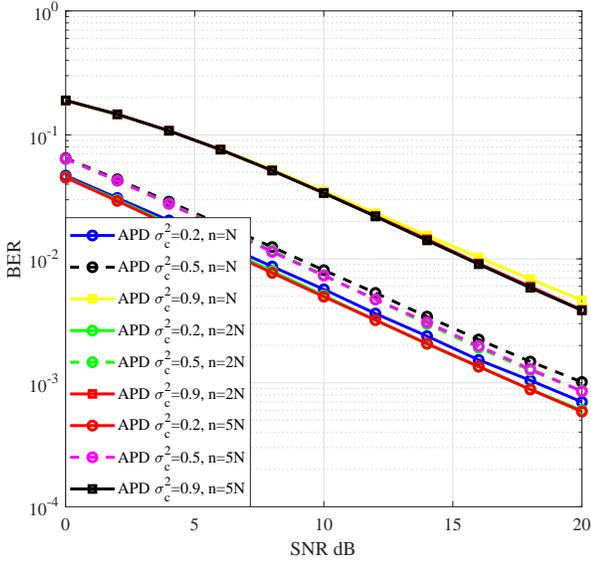}
  \caption{BER vs SNR of Proposed System for Different n. }
  \label{fig:Capture1}
\end{figure}
\begin{figure}[t!]
  \includegraphics[width=\linewidth]{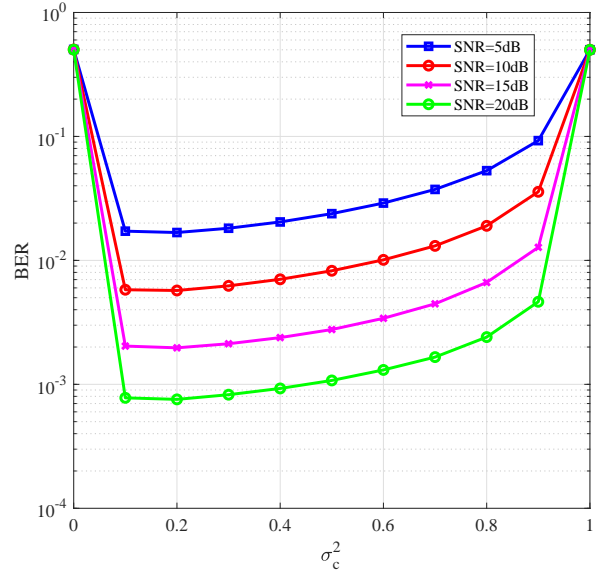}
  \caption{BER vs $\sigma_c^2$ of Proposed System $(n=N)$.}
  \label{fig:Capture4}
\end{figure}

  Fig.3 analyses   BER  performance of this  system for various value of $ n$. To satisfy the condition $n\geq N$ [7], we have taken the values of $n$ as $N, 2N, 5N $ and done the BER analysis. It is observed from Fig 3. that the proposed system BER performance improves as $n$ is increased. However, as $n $ value increases, the computational complexity of the system  increases and spectrum efficiency decreases (from ($\ref{zzz}$)).

Fig.4 represents  BER  of this system with $n=N$ for different $\sigma_c^2$ values. It is noticed that the BER performance is best for $\sigma_c^2=0.2$ for any value of SNR. Hence, affine precoding and decoding based FBMC system performs better than other FBMC systems and is best with $\sigma_c^2=0.2$.


\section{CONCLUSION}
We derived the mathemetical equations needed for channel estimation and data detection of SISO-FBMC/OQAM system with affine precoding and decoding. WIth the simulation results we proved that the BER of this method is better than other cutting edge methods.   The optimum training power coefficient $(\sigma_{c_{opt}}^2)$ is also found from  the simulation results. Though the proposed method is giving band width loss, from BER point of view the proposed FBMC system model is having practical importance.


\begin{thebibliography}{50}
\bibitem{3}	P. Siohan, C. Siclet, and N. Lacaille, "Analysis and design of
OQAM-OFDM systems based on filterbank theory," IEEE Trans. Signal
Process., vol. 50, no. 5, pp. 1170-1183, May 2002.
\bibitem{4} E. Kofidis and D. Katselis, "Preamble-based channel estimation in MIMO-OFDM/OQAM systems," in Proc. IEEE Int. Conf. Signal Image
Process. Appl., Nov. 2011, pp. 579-584.
\bibitem{5} E. Kofidis, D. Katselis, A. Rontogiannis, and S. Theodoridis, "Preamblebased channel estimation in OFDM/OQAM systems: A review," Signal
Process., vol. 93, no. 7, pp. 2038-2054, Jul. 2013.
\bibitem{6} Hu, S., Liu, Z., Guan, Y.L., Jin, C., Huang, Y. and Wu, J.M.,  "Training sequence design for efficient channel estimation in MIMO-FBMC systems". IEEE Access, 5, pp.4747-4758. 2017.
\bibitem{7} Kun Chen-Hu, Juan Carlos Estrada-Jimenez, M. Julia Fernandez-Getino Garca and Ana Garca Armada "Superimposed Training for Channel Estimation in FBMC-OQAM" Vehicular Technology Conference (VTC-Fall), 2017 IEEE 86th.
\bibitem{9}	M. Bellanger, "Specification and design of a prototype filter for filter bank based multicarrier transmission" in Proc. ICASSP, Salt Lake, UT, USA, May 2001, pp. 2417-2420.
\bibitem {8}N. N. Tran, D. H. Pham, H. D. Tuan and H. H. Nguyen, "Orthogonal Affine Precoding and Decoding for Channel Estimation and Source Detection in MIMO Frequency-Selective Fading Channels," in IEEE Transactions on Signal Processing, vol. 57, no. 3, pp. 1151-1162, March 2009, doi: 10.1109/TSP.2008.2009020.

\bibitem{10} P. Singh, R. Budhiraja and K. Vasudevan, "Probability of Error in MMSE Detection For MIMO-FBMC-OQAM Systems," in IEEE Transactions on Vehicular Technology, vol. 68, no. 8, pp. 8196-8200, Aug. 2019, doi: 10.1109/TVT.2019.2920475.
\end{thebibliography}
\end{document}